\documentclass[newversion]{aa}
\usepackage{graphicx}

\begin{document}


   \title{Observation of an Instability in a 'Quiescent' Prominence}

   \author{G. Stellmacher\inst{1}
          \and E. Wiehr\inst{2}}

   \offprints{E. Wiehr}

   \mail{ewiehr@astrophysik.uni-goettingen.de}

   \institute{Institute d'Astrophysique (IAP), 98 bis Blvd. d'Arago, 
              75014 Paris, France
              \and
              Institut f\"ur Astrophysik der Universit\"at,
              Friedrich-Hund-Platz 1, 37077 G\"ottingen, Germany}

   \date{Received Dec. 13, 1972; accepted March 8, 1973}

\abstract
{}
{We present the detection of a bubble-like cavity traveling through a 
quiescent prominence.}
{H$\alpha$ slit-yaw images were taken together with Ca\,II\,8542 spectra using 
an image intensifier.}
{The H$\alpha$ emission in the cavity is more than 16 times smaller than in its 
surroundings. The cavity propagates almost with the phase-velocity of MHD compressive 
waves. We suggest a disruption of the lateral magnetic stability criterion in the 
Kippenhahn-Schl\"uter (or Kuperus - Tandberg-Hanssen) model. The Ca\,II\,8542 spectra 
indicate a material outflow along the lines of force up to 12\,km/s.} 
{} 

\maketitle

%

\section{Observing Procedure}

The observation of prominence spectra with high spatial, spectral and time resolution 
is generally limited by the long exposure times required. For the \O = 45\,cm Locarno 
telescope with its 10\,m spectrograph ($\approx0.18$\,\AA/mm) the observation 
of the H$\alpha$ emission line in prominences would lead to an exposure time of several 
minutes. But the use of a new type image intensifier (two-step proximity-focused 'Bendix') 
attached to the focal plane of the spectrograph, reduces the exposure time to about 
one second. The image produced by the intensifier is taken on a Kodalith film at tight
contact to the fiber-optics output of the image intensifier. Its 24\,mm aperture permits 
observations of Doppler displacements up to $\pm120$\,km/s in H$\alpha$; its resolution 
of about 35\,lines/mm allows to spatially resolve 1\,arcsec (0.125 mm) and spectrally 
15\,m\AA{} (0.08\,mm for H$\alpha$; see also Stellmacher and Wiehr (1972).

%
%
	
\section{Description of the Event}

On October 12, 1971 we observed a quiescent prominence at the eastern limb (solar 
latitude $45^o$\,S) for a spectral analysis of various emission lines. The frequent 
configuration of an arch-like lower boundary of the prominence body unexpectedly 
ascended at 13.45\,UT. The series of H$\alpha$ slit-yaw-pictures (pass band 0.5 A)
indicates a projected motion between 7 and 17\,km/s (Fig.\,1). At 14.15 UT the lower 
part of the elevated arch closed up, thus forming a bubble-like 'cavity' within the 
prominence. This cavity continued to move upwards and contracted until it disappeared 
at 15.40 UT. The original overall configuration of the prominence was then restored 
and lasted the following day (Fig.\,3).
                     
%
   \begin{figure*}[ht]     
   \hspace{25mm}\includegraphics[width=13.5cm]{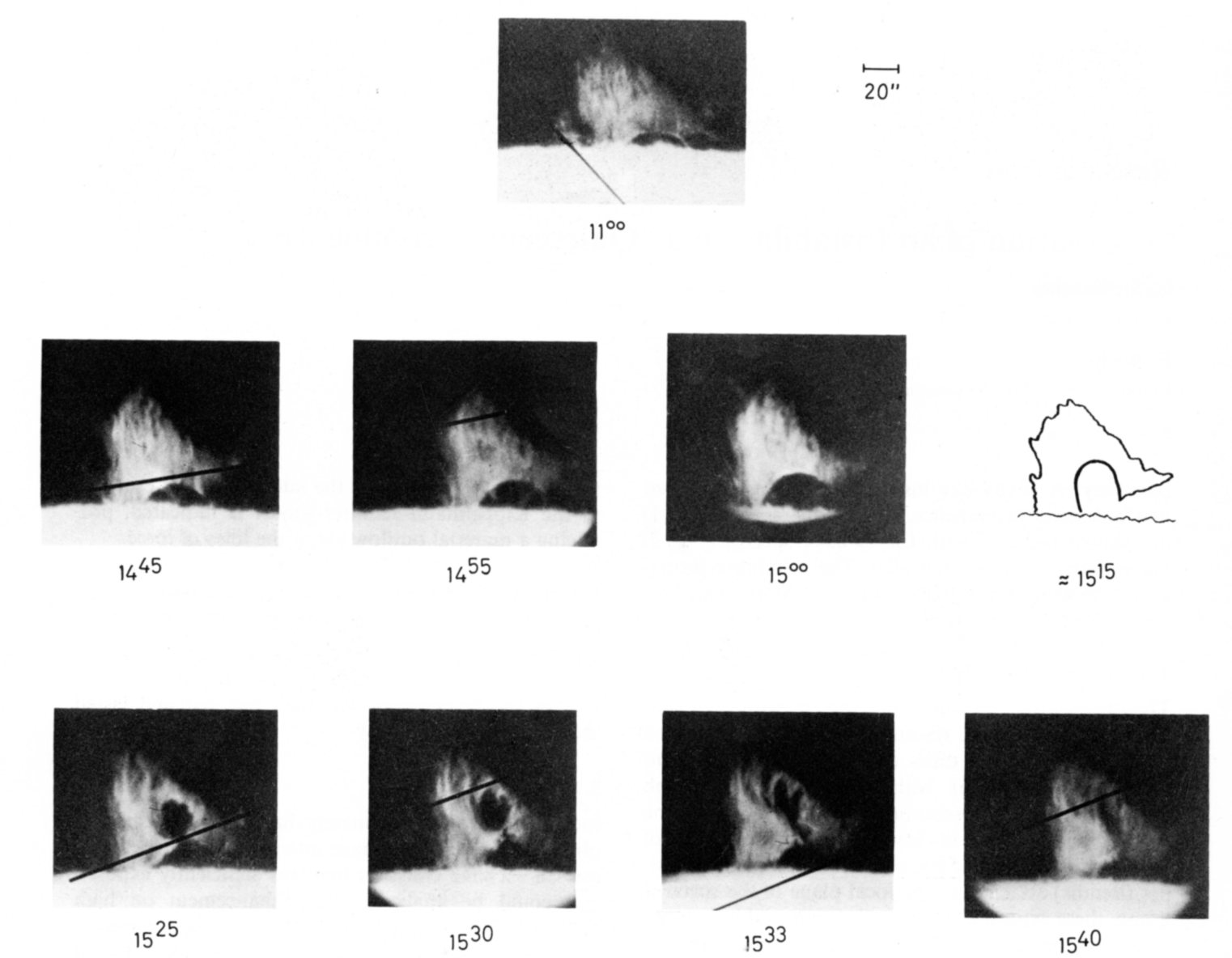}
   \caption{H$\alpha$ slit-yaw images of the 'cavity' event on Oct.\,12, 1971, 
in the 'quiescent' prominence at the east limb,$45^o$S. Due to problems with the 
slit-yaw camera, the 15.15 UT phase is given as a visual observation.}
   \label{Fig1}
    \end{figure*}
               
%

   \begin{figure*}[htb]     
   \hspace{25mm}\includegraphics[width=13.5cm]{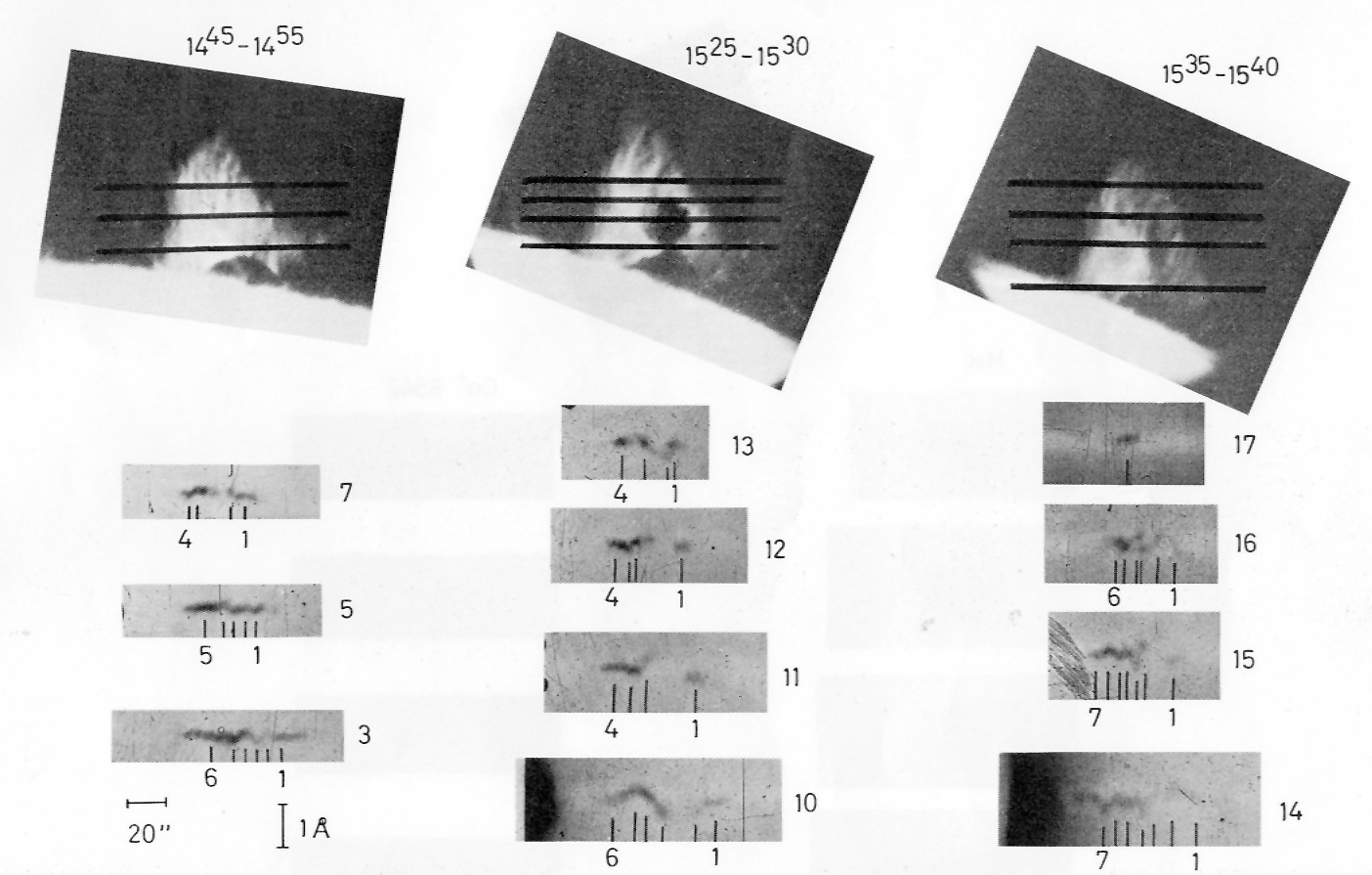}
   \caption{H$\alpha$ spectra of Ca\,II\,8542 taken during the cavity event; 
$\lambda$ increases to the top. The slit positions are entered in the corresponding 
H$\alpha$ images from Fig.\,1; scan locations are indicated in each spectrum.}
   \label{Fig2}
    \end{figure*}

%
%
	
\section{Spectra}
\subsection{Line shifts}

Increasing haze unfortunately hampered most of the spectral observations with the image 
intensifier. Only the infrared Ca\,II\,8542 emission line was sufficiently exposed and 
could thus be analyzed after substantial contrast enhancement of the Kodalith film 
(Fig.\,2). The prominence spectra taken before the formation of the cavity ascension 
(No.\,3-7 in Fig.\,2) show the characteristically wiggled lines, which equally occurred 
on the following day in both emissions Ca\,II\,8542 and H$\alpha$ (Fig.\,3). 

At the beginning of the cavity ascension, Doppler shifts of $v_r\approx3$\,km/s
occurred just above the lower boundary of the arch (points\,3 and 4 in spectrum\,3). 
During the ascension of the cavity, these shifts then occurred all over the prominence 
body, but increased at up to 13\,km/s at the boundary of the cavity (cf., locations 10/3 
and 13/2). They finally disappeared after the prominence became again stable at about 16\,UT.
             
\subsection{Line widths}

Measurements of the half-widths at 1/e-intensity, $\Delta\lambda_e$, are given in Table\,1 
together with the corresponding radial velocities, $v_r$. A correlation between both 
(as discussed by Engvold, 1972) could not be established. Most of the highly shifted line
profiles appear strongly broadened (10/3, 11/1, 11/2, 12/1, 13/2, 14/7), corresponding 
to non-thermal velocities of up to $v_{nth}\approx 5$\,km/s if $T_{kin} = 6000$\,K 
(Stellmacher, 1969). Some of these highly diffuse and asymmetric profiles show several 
peaks, which may indicate a superposition of different Doppler shifted emissions from 
individual prominence filaments with $v_{nth}= 1-2$\,km/s.

%
%
%

   \begin{figure*}[htb]     
   \hspace{30mm}\includegraphics[width=12.4cm]{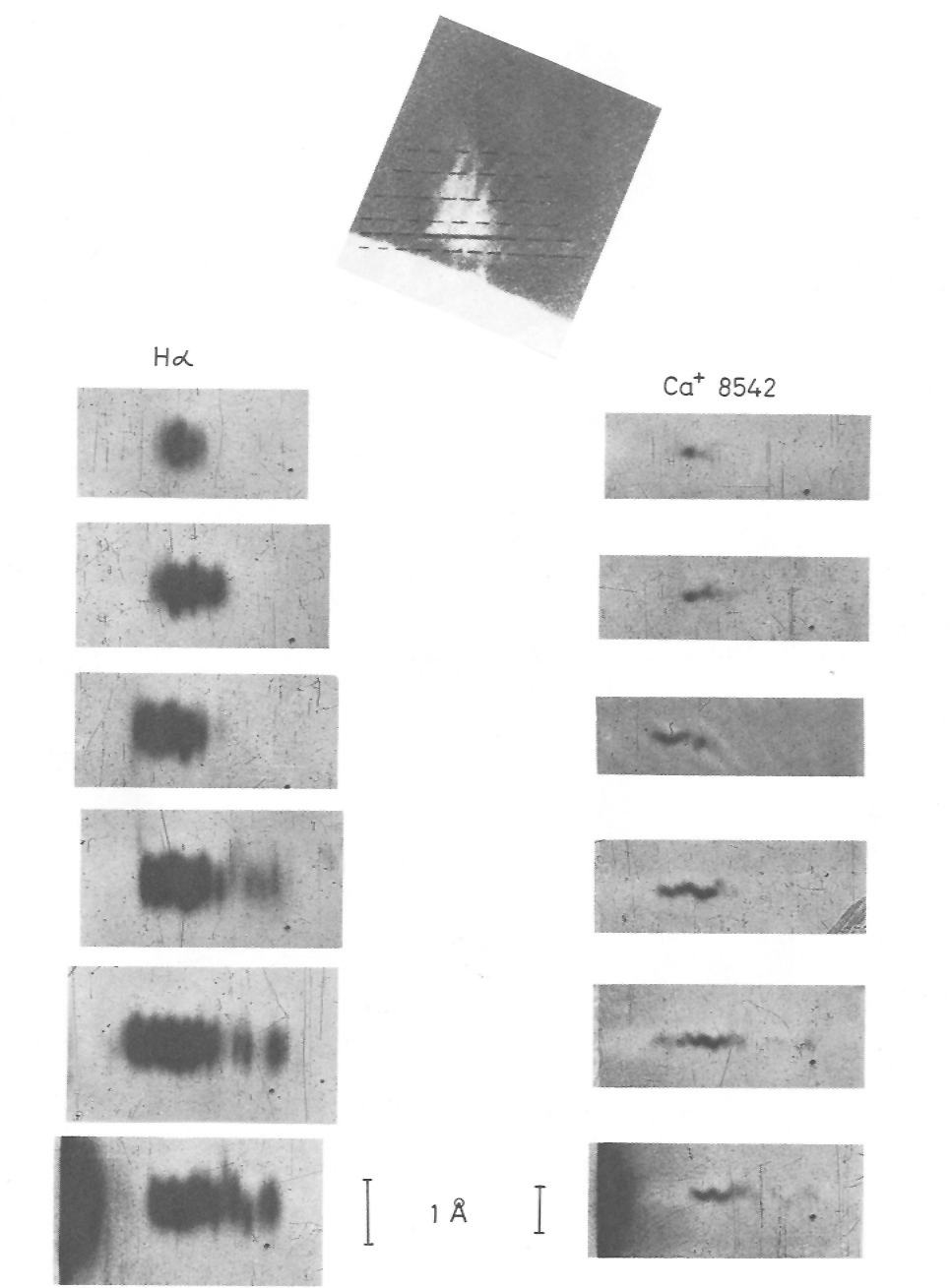}
   \caption{H$\alpha$ and Ca\,II\,8542 emissions in the same prominence taken the 
next day; $\lambda$ increases to the top.}
    \label{Fig3}
    \end{figure*}

\section{Emission inside the cavity}

From the Ca\,II\,8542 spectra 11, 12 and 15 it can be seen that the cavity 
represents a real disappearance of the emission and not a Doppler shift 'off-band' 
the H$\alpha$ filter of the slit-yaw imaging device. For an estimate of the residual 
H$\alpha$ emission inside the bubble-like 'cavity', we take as a lower limit for 
the detection of any prominence emission the aureole intensity of $3\cdot 10^{-3}$ 
of the continuum intensity at disk center. The residual emission inside the cavity
will than be smaller than 0.06 of the mean intensity of the prominence; the 
H$\alpha$ emission is thus reduced by at least a factor of 16 with respect to the 
surrounding emission of the undisturbed prominence body.
                      
%
%

\section{Discussion}

In the picture of recent prominence models (e.g. Kippenhahn and Schl\"uter, 1957, 
Kuperus and Tandberg-Hanssen, 1967; Anzer and Tandberg-Hanssen, 1970) the magnetic 
field lines within a quiescent prominence are almost parallel to the solar surface. 
{\it The bubble-like cavity would then have propagated more or less perpendicular 
to the lines of force.} It is therefore rather suggestive to consider MHD 
compressive waves as physical origin. Their phase velocity 
is $v_{MHD} = \sqrt{v^2_{sound} + v^2_{Alfven}}$ where $v_{sound}= 7$\,km/s if 
$T = 6000$\,K for the prominence. The observed 7\,-\,17\,km/s motion would then 
require an Alv\'en velocity of 3\,km/s$\le v_A\le13$\,km/s. This, in turn, corresponds 
to a (reasonable) strength of the prominence magnetic field of 0.5 Gs$\le B\le 1.5$\,Gs, 
if $q= 1.8\cdot 10^{-13}$\,g/cm$^3$ (in agreement with observed $N= 10^{11}$cm$^{-3}$).

The Doppler displacements at the 'cavity' border indicate a {\it lateral outflow of 
prominence matter, caused by the bubble-like disturbance}. It had already been 
suggested by Kippenhahn and Schl\"uter (1957) that relatively small disturbances 
of the supporting magnetic field may be sufficient to violate their stability criteria. 
In our case the stability criterion against lateral displacements may be disturbed, 
thus causing a lateral material outflow along the lines of force. The rather good 
agreement between the observed radial velocities (up to 12 km/s) and the sinking velocity 
of prominence matter through the corona (Uns\"old, 1970) seems to support this idea.

Further observations of such bubble-like instabilities in 'quiescent' prominences would 
be necessary to improve our knowledge about their physical origin and the stability 
criteria of prominences.
%
%

   \begin{table*}[ht]
   \caption[]{Doppler velocity, $v_r$, and half-width at 1/e intensity,
$\Delta\lambda_e$, of for each spatial location in the Ca\,II\,8542 
spectra Fig.\,2; brackets indicate estimates for location with too faint 
Ca\,II emission.}
   \hspace{-2mm}\includegraphics[width=19.0cm]{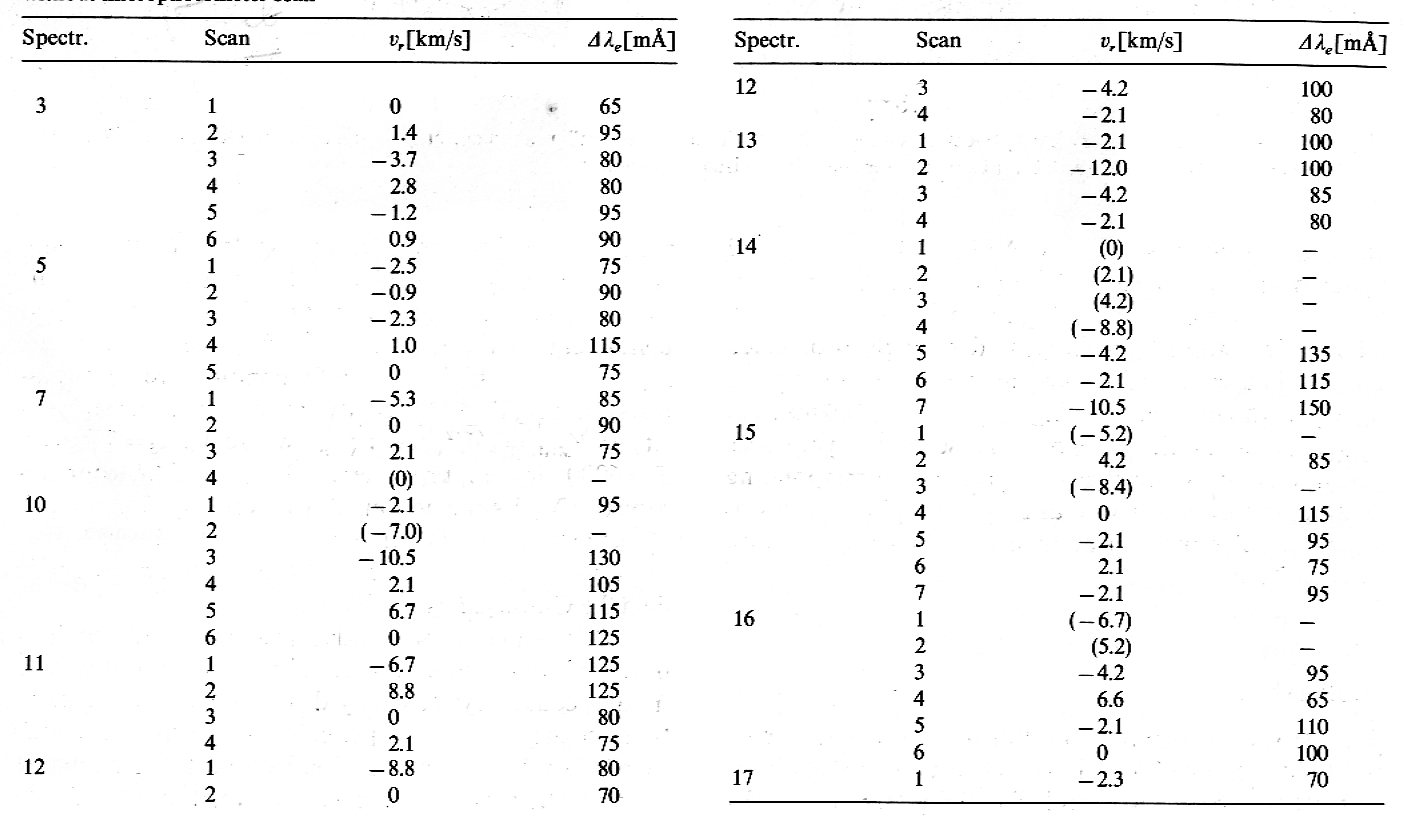}
   \label{Tab2}   
   \end{table*}

%
%

\begin{acknowledgements}
We want to thank R. Kippenhahn, I. Appenzeller and H. K\"ohler for interesting 
discussions. We are indebted to R. Spindler for his help with the photographic 
contrast enhancements of the spectra. The observations were carried out at the 
Locarno observatory of the 'Deutsche Forschungsgemeinschaft', operated by the
University of G\"ottingen.  
\end{acknowledgements}

\eject
%
%

%
%

\end{document}